\documentclass[twocolumn,preprintnumbers,amsmath,amssymb]{revtex4}
\usepackage{graphicx}
\usepackage{dcolumn}
\usepackage{bm}
\begin{document}
\title{
Effect of heating and cooling of photogenerated electron-hole plasma
in optically pumped graphene on population inversion. 
}
\author{V.~Ryzhii$^{1,4}$,  M.~Ryzhii$^{1,4}$, 
V.~Mitin$^{2}$,
A.~Satou$^{3,4}$, and   
 T.~Otsuji$^{3,4}$
}
\affiliation{
$^1$ Computational Nanoelectronics Laboratory, University of Aizu, 
Aizu-Wakamatsu, 965-8580, Japan\\
$^2$ Department of Electrical Engineering, University at Buffalo, Buffalo,
NY 1460-1920, U.S.A.\\
$^3$ Research Institute for Electrical Communication,
Tohoku University,  Sendai,  980-8577, Japan\\
$^4$ Japan Science and Technology Agency, CREST, Tokyo 107-0075, Japan\\
              }
\begin{abstract}{We study
 the characteristics of photogenerated electron-hole plasma in optically
pumped graphene layers  at elevated (room)  temperatures
when the interband and intraband processes of emission and absorption of optical phonons play a crucial role. 
The electron-hole plasma heating and cooling as well as the effect of nonequilibrium optical phonons are taken into account.
The dependences of the quasi-Fermi energy and  effective temperature of optically pumped graphene layers on the intensity of pumping
radiation are calculated. The variation of the frequency dependences
dynamic
conductivity with increasing pumping intensity  as well as the conditions
when this conductivity becomes negative in a certain range of frequencies
are considered.
 The effects under consideration  can markedly influence the achievement of
the negative dynamic conductivity in optically pumped
graphene layers  associated with the population inversion and, hence, 
the realization graphene-based terahertz and infrared lasers operating  at room temperatures. 
}
\end{abstract}
\maketitle

\section{Introduction}

The gapless energy spectrum of graphene layers (GLs)~\cite{1} provides an opportunity
to create devices utilizing the interband transitions
in a wide spectral range of radiation.
As shown~\cite{2} (see also Refs.~\cite{3,4,5}),
under optical or electrical pumping, the interband population inversion can 
be realized in GLs. Due to fairly high quantum efficiency of the interband
transitions in GL~\cite{6,7} and particularly in multiple-GL structures~\cite{8,9}, 
the interband  contribution to 
the real part of the dynamic conductivity  can surpass that associated with the intraband (Drude) absorption~\cite{2}, so that the GL net dynamic conductivity is negative in a certain range of signal frequencies.
This opens up the prospects  of achieving of lasing of 
the terahertz (THz) and 
infrared (IR) radiation  in different
structures on the base of single- and multiple-GLs~\cite{10,11,12}.
In optically pumped GLs,
a  marked portion of the absorbed optical
energy   can go to the electron and hole 
heating.
As demonstrated~\cite{5,13,14,15}, the electron and hole heating hinders 
the realization of population inversion and negativity of  dynamic conductivity. 
At elevated (room)  temperatures, in contrast to the
low temperature situation~\cite{13},
the interband transitions associated with optical phonons can be the dominant mechanism of the 
recombination~\cite{16}.  
The energy relaxation of electrons and holes in GLs can also be mainly due to
interaction with optical phonons~\cite{17}. In such a case,
the optical photons emitted by the photogenerated electrons and holes
can accumulate in GL, so that the optical phonon system is heated as well. 
This can lead to the optical photons reabsorption accompanied with the intraband
electron and hole transitions. The deviation of the optical
phonon system from equilibrium 
 can also affect the properties of GL, as it occurs in carbon nanotubes
(see for instance Refs.~\cite{18,19,20,21,22,23,24}),
in particular, affect
the interband generation-recombination processes.
Increase of the number 
of nonequilibrium phonons result in the increase of stimulated optical 
phonon emission and may lead to the generation of
coherent phonons in graphene like it was 
discussed for a different systems~\cite{25,26,27}.
Thus, the accounting for the optical phonon heating appears to 
be indispensable.
However, it is important to stress that at certain values of the pumping radiation frequencies,
cooling of the electron-hole plasma below the temperature of the lattice 
is possible.

In this paper, we study the effect of  heating and cooling 
of the electron-hole plasma in optically pumped GLs at elevated temperatures,
when the interaction with optical phonons is the main mechanism of the recombination and energy relaxation,  
on its  characteristics, particularly on  its dynamic conductivity
and in the THz and IR frequency ranges. 
The deviation of the optical phonon system from equilibrium (optical phonon
heating) is taken into account
The obtained characteristics are  crucial for the  
realization of THz and IR lasers~\cite{2} as well as different THz 
plasma wave
devices~\cite{28,29} on  
GL structures operating at 
room temperatures and can be used for the device optimization.

\section{Model and the pertinent equations}

We consider the interband phogeneration of electrons and holes by optical radiation with the photon energy $\hbar\Omega$ (optical pumping), where $\hbar$ 
is the reduced Planck constant.
After the photogeneration of an electron and a hole with the energy exceeding
the optical phonon energy$\hbar\omega_0 \sim 200$~meV, their behavior 
can correspond to the following tracks:\\
(a) If the characteristic time, $\tau_{cc}$, of inter-carrier scattering (electron-electron, electron-hole, and hole-hole) is much 
larger than the characteristic time, $\tau_0$ of 
spontaneous emission of  
optical phonon    just photogenerated carriers can manage to emit 
cascades of optical phonons. The number of  optical  phonons $K$ in such a cascade
is determined by the ration $\Omega/\omega_0$.
Due to the symmetry of the energy spectrum, 
the photogenerated electrons and holes
emit equal number ($K$) of phonons. 
This cascade is followed by the electron and hole "fermisation" 
and recombination~\cite{2}, so that the electron and hole distribution function
becomes equal to  $f = 
\{\exp[(\varepsilon - \varepsilon_F)/T] + 1\}^{-1}$. Here
$\varepsilon = v_Wp$ is the energy of electrons and  holes
 with the momentum $p$,
$v_W = 10^8$~cm/s is the characteristic velocity of the GL energy spectrum,
$\varepsilon_F$ is the quasi-Fermi energy, and $T$ is the effective temperature
(in the energy units). The quasi Fermi energy of the electron-hole
plasma in the situation in question
 $\varepsilon_F$ and its effective temperature
$T$ are determined by the rates of photogeneration ,and 
 recombination,  as well as
the rates intraband and interband energy relaxation.
These parameters are determined by the energy $\varepsilon_0$
which is supplied to the electron-hole plasma in each act of the photogeneration. 
In the case in question,   $\varepsilon_0 =\hbar\Omega_0
= \hbar\Omega - 2K\hbar\omega_0$.
Generally, $\varepsilon_F$ and 
$T$ are not equal to their equilibrium values, i.e.,  $\varepsilon_F \neq 0$
and $T \neq T_0$;\\
(b) If the  inter-carrier scattering time is the shortest one 
($\tau_{cc} \ll \tau_0$),
just photogenerated electrons and holes are immediately
"fermilised". In this case, all the energy of  photogenerated carriers goes
to the electron-hole plasma, so that $\varepsilon_0 = \hbar\Omega$;\\
(c) At relatively low energies of photons ($\hbar\Omega < 2\hbar\omega_0$),
the situation is similar to the previous case 
 with  $\varepsilon_0 = \hbar\Omega$, 
but for rather arbitrary relationships
between the characteristic times of inter-carrier and optical phonon scattering.
This case corresponds, in particular,
 to optical pumping by CO$_2$ or quantum-cascade lasers;\\
(d) When $\hbar\Omega > \hbar\omega_0$ (or  $\hbar\Omega \gg \hbar\omega_0$),
and  the characteristic times of inter-carrier and optical phonon scattering
are of the same order of magnitude ($tau_{cc} \sim \tau_0$), 
the pattern of the energy relaxation
of photogenerated carriers can be fairly complex. 
Nevertheless, even in this case, on can introduce the energy 
$\varepsilon_0 = \hbar\Omega_0$,
which, however, is determined not only by  the ration of 
$\hbar\Omega$ and $\hbar\omega_0$, but also by the ratio of the characteristic
scattering times.  
Naturally, $\hbar\Omega - 2K\hbar\omega_0 < \varepsilon_0 < \hbar\Omega$. In this case,
the quantity $\varepsilon_0$ (or $\Omega_0$) is a phenomenological parameter,
which determines the fractions of the absorbed photon energy going to
directly to the electron-hole plasma and to the optical phonon system.
For a rough estimate,  the following simple formula  might be used:
$\hbar\Omega_0 = \hbar\Omega - 2K\hbar\omega_0/[1 + K\tau_0/\tau_{cc}]$,
which yields the above values of $\hbar\Omega_0$ in the limiting cases
(a) and (b).

Introducing parameter  $\varepsilon_0$ (or $\Omega_0$), one can
consider all the abovementioned tracks in the framework of the same treatment.

In the situation under consideration, the optical phonon
system can also be far from equilibrium, so that the distribution function
of optical phonons ${\cal N}_0$ can markedly deviated from
its equilibrium value ${\cal N}_{0}^{eq} = [\exp(\hbar\omega_0/T_0) - 1]^{-1}$.
The latter can lead to an effective reabsorption of optical phonons
by electrons and holes with a significant increase in their energies.
Although at low temperatures, the energy relaxation and recombination
 of electrons and holes 
are  associated with the interaction with acoustic phonons and 
radiative processes, respectively~\cite{13}, in the case of elevated temperatures
(say, room temperatures)
 and pumping intensities, the energy relaxation and recombination
is assumed to be due to the interactions with optical phonons (both inraband and interband). 

The quasi-Fermi energy of the electron-hole plasma $\varepsilon_F$,
 its effective temperature $T$, and the number of optical phonons ${\cal N}_0$
obey the equations

\begin{equation}\label{eq1}
 R_0^{inter}  = G_{\Omega},
\end{equation}
\begin{equation}\label{eq2}
\hbar\omega_0(R_0^{inter}
+ R_0^{intra})   = 
G_{\Omega_0}
\hbar\Omega_0,
\end{equation} 
\begin{equation}\label{eq3}
 \hbar\omega_0\, R^{decay} = \hbar\Omega\,G_{\Omega},
\end{equation}
governing
the balance of the electron-hole pairs,  the electron-hole plasma 
energy balance, respectively, as well as  the optical phonon  balance. 
Here, $R_0^{inter}$,  $R_0^{intra}$,
and $R^{decay}$ are the rates
of the pertinent processes.
Equation~(3)  explicitly takes into account that all the energy
received by the system from optical pumping goes eventually to the thermostat
(to the contacts via acoustic phonons).  
 The optical generation rate is given by
\begin{equation}\label{eq4}
G_{\Omega} = \pi\alpha\tanh\biggl(\frac{\hbar\Omega - 2\varepsilon_F}{4T}\biggr)\,I.
\end{equation}
Here, 
$I$ is the photon flux of pumping radiation and 
$\alpha \simeq 1/137$ is the fine structure constant 
(so that the absorption coefficient is equal to $\pi\alpha \simeq 0.023$).
The last factor in the right-hand side of Eq.~(4) accounts for
the limitation of the interband absorption associated with the Pauli blocking.
In this case, 
$G_{\Omega} \simeq G_{\Omega_0} \simeq \pi\alpha\,I$.  
To provide an effective cascade pumping and 
avoid the fast direct
transitions with the emission of an additional  optical phonon, the following conditions should be satisfied: 
\begin{equation}\label{eq5}
\varepsilon_F < \hbar\Omega_0/2 <  \hbar\omega_0 - \varepsilon_F
\end{equation}

\section{Rates of the processes under consideration}

 For the terms $R_0^{inter}$ and $R_0^{intra}$ , which describe the
electron-hole recombination and generation processes 
and the intraband energy relaxation
assisted by optical phonons,
 one can use the following simplified
formulas:

\begin{equation}\label{eq6}
R_0^{inter} = \frac{\Sigma_0}{\tau_0^{inter}}
\biggl[
({\cal N}_0 + 1)\exp\biggl(\frac{2\varepsilon_F - \hbar\omega_0}{T}\biggr)  - {\cal N}_0
\biggr],
\end{equation}
\begin{equation}\label{eq7}
R_0^{intra} =  \frac{\Sigma_0}{\tau_0^{intra}}
\biggl[({\cal N}_0 + 1)\exp\biggl(-\frac{\hbar\omega_0}{T}\biggr) 
- {\cal N}_0\biggr]. 
\end{equation}
Here, 
$\Sigma_0 = \pi(T_0/\hbar\,v_W)^2/6$ 
is the equilibrium electron and hole density
and $\tau_0^{inter} \propto \tau_0^{intra} \propto \tau_0$  are the times 
of the  interband and intraband  
phonon-assisted processes. The latter quantities have been introduced
in the present form for convenience

A difference in  $\tau_0^{inter}$ and  $\tau_0^{intra}$ is associated
with the features of the density of states and with  a difference between the optical phonon energy and the quasi-Fermi energy (see Appendix).
In equilibrium, i.e., at ${\cal N}_{0} = {\cal N}_{0}^{eq}$, $\varepsilon_F = 0$, 
and $T = T_0$, Eqs.~(6) and (7) yield $R_0^{inter} = R_0^{intra} = 0$.

The rate of optical phonons decay due to the anharmonic contributions to the
interatomic potential, resulting in the phonon-phonon
scattering and in  the decay of optical phonons into
acoustic phonons and 
is assumed to be in the following form:

\begin{equation}\label{eq8}
R_0^{decay} = \frac{\Sigma_0({\cal N}_0 - {\cal N}_{0}^{eq}) }{\tau_0^{decay}},
\end{equation}
where $\tau_0^{decay}$ is the pertinent characteristic time.
Considering high heat conductivity of GLs~\cite{30},
the lattice  temperature, i.e. the temperature
 of acoustic phonons,  is assumed to be equal to 
the temperature of the contact $T_0$.

\section{Calculation of the effective temperature and quasi-Fermi energy}

Equations~(1), (3), and (8) yield
\begin{equation}\label{eq9}
{\cal N}_0 =  {\cal N}_{0}^{eq}\biggl(1  +
\eta^{decay}_0\frac{I}{I_0}\biggr).
\end{equation}
Here 
$$
I_0 = {\cal N}_{0}^{eq}\frac{\omega_0}{\Omega}\frac{\Sigma_0}{\pi\alpha\tau_0^{inter}}
\simeq \exp\biggl(- \frac{\hbar\omega_0}{T_0}\biggr)\frac{\omega_0}{\Omega}\frac{\Sigma_0}{\pi\alpha\tau_0^{inter}}
$$
is  the characteristic photon flux and $\eta^{decay}_0 = 
\tau_0^{decay}/\tau_0^{inter}$.
As follows from Eq.~(9), at very large pumping intensities,
${\cal N}_0$ can markedly exceed unity. We shall not consider the range of such intensities. Therefore, in the following we limit our consideration 
by not too strong pumping assuming that
$\eta_0^{decay} I/I_0 \ll  \exp(\hbar\omega_0/T_0)$, i.e., $I \ll 
I_0\exp(\hbar\omega_0/T_0)/\eta_0^{decay} = I_0^{\infty}$.

Using Eqs.~(1), (2), (5), and (7) with Eq.~(9), 
  we arrive at the following equations for the effective temperature $T$
and  the quasi-Fermi energy
$\varepsilon_F$ (for not too strong pumping intensities):

$$
\exp\biggl(\frac{2\varepsilon_F - \hbar\omega_0}{k_BT}\biggr) 
=
\frac{1 + 
 \displaystyle(\eta_0^{decay} + 1)
\frac{I}{I_0}}
{\displaystyle\exp\biggl(\frac{\hbar\omega_0}{T_0}\biggr) + 1 + \displaystyle\eta_0^{decay}\frac{I}{I_0}}
$$
\begin{equation}\label{eq10}
\simeq \exp\biggl(- \frac{\hbar\omega_0}{T_0}\biggr)\biggr[1 + 
 \displaystyle(\eta_0^{decay} + 1)
\frac{I}{I_0}\biggr],
\end{equation}

$$
\exp\biggl(-\frac{\hbar\omega_0}{k_BT}\biggr) 
=
\frac{1 + 
\displaystyle\biggl[\eta_0^{decay} + 
\eta_0\biggl(\frac{\Omega_0}{\omega_0} - 1\biggr)\biggr]\frac{I}{I_0}}
{\displaystyle\exp\biggl(\frac{\hbar\omega_0}{T_0}\biggr) +  1 +
\displaystyle\eta_0^{decay}\frac{I}{I_0}}
$$
\begin{equation}\label{eq11}
\simeq \exp\biggl(- \frac{\hbar\omega_0}{T_0}\biggr)\biggl\{1 + 
\displaystyle\biggl[\eta_0^{decay} + 
\eta_0\biggl(\frac{\Omega_0}{\omega_0} - 1\biggr)\biggr]\frac{I}{I_0}\biggr\}
,
\end{equation}
\begin{equation}\label{eq12}
\exp\biggl(\frac{2\varepsilon_F }{T}\biggr) =
\frac{1 + 
\displaystyle (\eta_0^{decay} + 1)
\frac{I}{I_0}}
{1 + 
\displaystyle\biggl[\eta_0^{decay} +  
\eta_0\biggl(\frac{\Omega_0}{\omega_0} - 1\biggr)\biggr]\frac{I}{I_0}}.
\end{equation}
Here $\eta_0 = \tau_0^{intra}/\tau_0^{inter}$.

Equations~(11) and (12) yield

\begin{widetext}
\begin{equation}\label{eq13}
T \simeq \frac{T_0}{1 - \displaystyle\frac{T_0}{\hbar\omega_0}\,
\ln\biggl\{1 +
\biggl[\eta_0^{decay} + \eta_0\biggl(\frac{\Omega_0}{\omega_0} - 1\biggr)
\biggr]\frac{I}{I_0}\biggr\}},
\end{equation}

\begin{equation}\label{eq14}
\varepsilon_F \simeq \frac{T_0}{2}\,
\frac{\ln \biggl\{\frac{\displaystyle 1 + 
\displaystyle (\eta_0^{decay} + 1)
\frac{I}{I_0}}
{\displaystyle 1 + 
\displaystyle\biggl[\eta_0^{decay} +  
\eta_0\biggl(\frac{\Omega_0}{\omega_0} - 1\biggr)\biggr]
\frac{I}{I_0}}\biggr\}}{1  - \displaystyle\frac{T_0}{\hbar\omega_0}\,
\ln\biggl\{1 +
\biggl[\eta_0^{decay} + \eta_0\biggl(\frac{\Omega_0}{\omega_0} - 1\biggr)
\biggr]\frac{I}{I_0}\biggr\}}.
\end{equation}
\end{widetext}
As it should be, Eqs.~(13) and (14) yield $T = T_0$ and $\varepsilon_F = 0$
if $I = 0$. 

The dependence of parameter $\eta_0$ on $\varepsilon_F$ and $T$
complicates the solution of Eqs.~(13) and (14) 
in wide ranges of pumping intensities, particularly, their
analytical solution. 
  
\section{Special and limiting cases}

\subsection{ Special case  $\Omega_0/\omega_0 = 1$.}

In a special case $\Omega_0/\omega_0 = 1$ (i.e., $\Omega = (2K + 1)\omega_0$), 
Eqs.~(12) - (14) yield

$$
T \simeq \frac{T_0}{ \displaystyle 1 -\frac{T_0}{\hbar\omega_0}\,
\ln(1 + \eta_0^{decay}I/I_0)}
$$ 
\begin{equation}\label{eq15}
\simeq T_0
+ \displaystyle\frac{T_0^2}{\hbar\omega_0}\,
\ln(1 + \eta_0^{decay}I/I_0),
\end{equation}

$$
\varepsilon_F \simeq \frac{T_0}{2}\,
\frac{\ln \biggl(1 + \frac{\displaystyle I/I_0}
{\displaystyle 1 + 
\displaystyle\eta_0^{decay}I/I_0}\biggr)}
{1  - \displaystyle\frac{T_0}{\hbar\omega_0}\,
\ln(1 +
\eta_0^{decay}I/I_0)} 
$$
\begin{equation}\label{eq16}
\simeq \frac{T_0}{2}\,
\ln\biggl(1 + \frac{I/I_0}
{\displaystyle 1 + 
\displaystyle\eta_0^{decay}I/I_0}\biggr),
\end{equation}
\begin{equation}\label{eq17}
\exp\biggl(\frac{2\varepsilon_F }{T}\biggr) 
= 1 + \frac{\displaystyle I/I_0}
{\displaystyle 1 + 
\displaystyle\eta_0^{decay}I/I_0}.
\end{equation}
Thus, at $\Omega_0/\omega_0 = 1$, the effective temperature $T$ 
is an increasing (logarithmic)  function of the pumping intensity $I$, 
although the increase in $T$ is rather slow due to a small factor $T_0/\hbar\omega_0 \simeq 0.125$
in the last pre-logarithmic term in the right-hand side of Eq.~(15).
As follows from Eq.~(16), the quasi-Fermi
energy $\varepsilon_F$ logarithmically  increases with $I$ and tends to 
the saturates with 
 $\varepsilon_F^{sat} = (T_0/2)\ln(1 + 1/\eta_0^{decay})$.
The increase in $\varepsilon_F$ in the range of moderate $I/I_0$
($I/I_0 \lesssim 1/\eta_{decay}$) is faster than that in $T$.
One can also see that an increase in the decay time leads to
slowing down the $\varepsilon_F - I$ dependence.

One can see that the variation of the effective temperature (the electron-hole plasma heating) is associate with the finiteness of the optical phonon decay 
time, i.e., with  the optical phonon heating and the absorption of hot phonons by electrons and holes, while 
the contributions of the
intraband and interband emission and absorption of optical phonons to 
the electron-hole plasma energy balance at $\Omega_0/\omega_0$ are compensated.
It is instructive that in the case in question, the effective temperature
of the optical phonon system $\Theta$, defined in  such a way that ${\cal N}_0 = [\exp(\hbar\omega_0/\Theta) - 1]^{-1}$, exceeds the lattice temperature $T_0$
(optical phonon heating) and
is equal to the effective temperature of 
the electron-hole plasma $T$. 
Indeed, using Eq.~(9), we find (compare with Eq.~(15))
 \begin{equation}\label{eq18}
\Theta \simeq \frac{T_0}{ \displaystyle 1 -\frac{T_0}{\hbar\omega_0}\,
\ln(1 + \eta_0^{decay}I/I_0)}.
\end{equation}
The fact that $\Theta =  T > T_0$ implies that both electron-hole plasma and
optical phonon system are heated and they are in equilibrium with each other.

If $\hbar\Omega = 1$~eV, so that $K = 2$, assuming that $T_0 = 300$~K,
and $\eta_0^{decay}
= 1 - 3$, and setting $I/I_0 =  1/\eta_0^{decay}$,
we obtain $T = \Theta \simeq 326$K,
When
$I/I_0 > 1/\eta_0^{decay}$, we obtain
 $\varepsilon_F \simeq \varepsilon_F^{sat}\simeq  (7 - 17)$~meV, 
and $\exp(\varepsilon_F^{sat}/2T) \lesssim 1.07  - 1.19$.

The optical phonon decay time and, hence, parameter $\eta_{0}^{decay}$
might be small in GLs on properly chosen  substrate.
In such a case, the quasi-Fermi energy at sufficiently strong pumping
can be not so small. 
Indeed, setting  $I/I_0 = 1/\eta_0^{decay}$ and 
$\eta_0^{decay}
= 0.1 - 0.5$, from Eqs.~(15) - (17)
we obtain $\varepsilon_F^{sat}\lesssim 17 -45$~meV,
and $\exp(\varepsilon_F^{sat}/2T) \lesssim 1.19 - 1.57$.

The optical phonon decay time, particularly in  suspended  GLs, 
can be  fairly long (in the range $1 - 10$~ps~
\cite{21,22,23,31,32,33}),  and  it might markedly exceed the
characteristic time of intraband interaction of electrons and holes
with optical phonons. Hence, the inequality 
 $\eta_0^{decay} \gg \eta_0 > 1$ can,  be valid, particularly, in the case of suspended GLs. In this fairly realistic case,
Eqs.~(12) - (14) lead to Eqs.~(14) - (17) and their consequences
at all values of 
$\Omega_0/\omega_0$, because
the terms with $\eta_0 (\Omega_0/\omega_0 - 1)$ are relatively small and
can  be omitted.

Thus,  at long  optical phonon  decay times,
the effect of accumulation of optical phonons (heating of the
optical phonon system) can prevent realization of  population
inversion necessary for lasing
 (large quasi-Fermi energy $\varepsilon_F$) at room temperatures
when the processes involving optical phonons can dominate.

\subsection{Weak pumping}

At sufficiently weak pumping (see below),
Eqs.~(12) - (14) can be solved analytically at arbitrary values of 
$\eta_0^{decay}$,
$\eta_0$, and $\Omega_0/\omega_0$.
In this case,
the effective temperature  and quasi-Fermi energy
are  close to $T_0$ and zero, respectively.
As shown in the Appendix, at $\varepsilon_F < T$,
one obtains $\eta_0 = \eta_{0}^{eq} \simeq
(\hbar\omega_0/\pi\,T_0)^2/(1 + 2.19T_0/\hbar\omega_0)$.
At room temperature, $\eta_{0}^{eq} \simeq 5$.
Thus, at a weak pumping,
$$
T \simeq   T_0 + \displaystyle\frac{T_0^2}{\hbar\omega_0}
\displaystyle\biggl[\eta_0^{decay}
+ \eta_{0}^{eq}\biggl(\frac{\Omega_0}{\omega_0} - 1\biggr)\biggr]\frac{I}{I_0}
$$
\begin{equation}\label{eq19}
\simeq T_0 + \displaystyle\frac{T_0^2}{\hbar\omega_0}
\displaystyle\biggl[\eta_0^{decay}
+ 5\biggl(\frac{\Omega_0}{\omega_0} - 1\biggr)\biggr]\frac{I}{I_0},
\end{equation}
$$
\varepsilon_F \simeq \frac{T_0}{2}\biggl[1 -  
\eta_{0}^{eq}\biggl(\frac{\Omega_0}{\omega_0} - 1\biggr)\biggr]\,\frac{I}{I_0}
$$
\begin{equation}\label{eq20}
\simeq 3T_0\biggl(1 -  
\frac{5}{6}\frac{\Omega_0}{\omega_0}\biggr)\,\frac{I}{I_0}.
\end{equation}
Simultaneously, one obtains
$$
\exp\biggl(\frac{2\varepsilon_F }{T}\biggr) \simeq 
1 + \biggl[1  - \eta_{0}^{eq}\biggl(
\frac{\Omega_0}{\omega_0} - 1\biggr)
\biggr]\,\frac{I}{I_0} 
$$
\begin{equation}\label{eq21}
\simeq 1 +  3\biggl(1 -  
\frac{5}{6}\frac{\Omega_0}{\omega_0}\biggr)\,\frac{I}{I_0}.
\end{equation}
Equations (19) - (21) are valid if
$$
I \ll \frac{I_0}{\biggl|\eta_0^{decay}
+ \displaystyle\eta_{0}^{eq}\biggl(\frac{\Omega_0}{\omega_0} - 1\biggr)\biggr|}
= \frac{I_0}{|\eta^{eq}|},
$$
where
$\eta^{eq} =\eta_0^{decay} + \eta_{0}^{eq}(\Omega_0/\omega_0 - 1)$.

As clearly  seen from Eqs.~(19) and (20) (as well as from Eqs.~(13) and (14)),
 the effective temperature increases (the electron and hole heating)
with  increasing $I$ if $\eta^{eq} > 0$, i.e., if
$\Omega_0/\omega_0 > 1 - \eta_0^{decay}/\eta_{0}^{eq}$.
The rise of the effective temperature is more pronounced
at larger $\eta_0^{decay}$, i.e., at longer decay time  $\tau_0^{decay}$.
Under the  condition 
$\Omega_0/\omega_0 > 1 + 1/\eta_{0}^{eq} \simeq 6/5$, 
the quasi-Fermi energy can become negative $\varepsilon_F < 0$, so that
$\exp (\varepsilon_F/T) < 1$.
In the latter case, the optical pumping does not lead to the degeneration
of the electron-hole plasma. It is interesting that this conclusion
is independent of the relative value of the decay time $\tau_0^{decay}$.

However, at sufficiently small ratio $\Omega_0/\omega_0$, parameter $\eta^{eq}$ 
can be negative,
and  the effective temperature
can decrease (cooling of the electron-hole plasma).
If $\Omega_0/\omega_0 < 6/5$,
 the quasi-Fermi energy
is a rising function of the pumping intensity.
The latter inequality implies that there should be  
$\Omega < \omega_0(2K + 1.2)$.
The latter conditions of the electron-hole plasma cooling accompanied with
the quasi-Fermi energy increase do not contradict inequalities Eq.~(5).
The effect of cooling under consideration is attributed to 
the effective photogeneration of low-energy electrons and holes (with the energy 
$\hbar\Omega_0/2$)

\subsection{Strong pumping, 
$\Omega_0/\omega_0 > 1 - \eta_0^{eq}/\eta_0^{decay}$.}

In this case,
$\eta_0^{decay} + \eta_0^{eq}(\Omega_0/\omega_0 - 1) > 0$, even
at  elevated pumping intensities $I \gtrsim I_0$,  
a moderate  increase in  $T$ and $|\varepsilon_F|$ exhibit  a moderate  increase 
 with increasing $I/I_0$, which
 slows down in the range of large  $I/I_0$
(the $T - I$ dependence becomes logarithmic, whereas the quasi-Fermi energy
tends to a constant value, positive or negative). 
Indeed, in the case under consideration,
Eq.~(14) does not have solutions with $\varepsilon_F \gg T$.
Hence in the logarithm in Eq.~(14) one can put $\eta_0 = \eta_0^{eq} \simeq 5$.
As a result, we obtain
$$
T  \simeq T_0 + \displaystyle\biggl(\frac{T_0^2}{\hbar\omega_0}\biggr)
\ln\biggl\{
\displaystyle\biggl[\eta_0^{decay} + \eta_0^{eq}\biggl(\frac{\Omega_0}{\omega_0} - 1\biggr)\biggr]\frac{I}{I_0}\biggr\}
$$
\begin{equation}\label{eq22}
\simeq  T_0 + \displaystyle\biggl(\frac{T_0^2}{\hbar\omega_0}\biggr)
\ln\biggl\{
\displaystyle\biggl[\eta_0^{decay} + 5\biggl(\frac{\Omega_0}{\omega_0} - 1\biggr)\biggr]\frac{I}{I_0}\biggr\},
\end{equation}

$$
\varepsilon_F \simeq \frac{T_0}{2}\,
\ln \biggl[\frac{\eta_0^{decay} + 1}
{\eta_0^{decay} +  \displaystyle
\eta_0^{eq}\biggl(\frac{\Omega_0}{\omega_0} - 1\biggr)}\biggr]
$$
\begin{equation}\label{eq23}
\simeq \frac{T_0}{2}\,
\ln \biggl[\frac{\eta_0^{decay} + 1}
{\eta_0^{decay} +  \displaystyle
5\biggl(\frac{\Omega_0}{\omega_0} - 1\biggr)}\biggr] = \varepsilon_F^{sat}.
\end{equation}

When $\eta_0^{decay} \gg 1$, Eq.~(23) yields
$$
\varepsilon_F^{sat} \simeq \frac{T_0}{2}\,
\ln \biggl[1 + \frac{5}{\eta_0^{decay}}
\biggl(\frac{6}{5} - \frac{\Omega_0}{\omega_0}\biggr)\biggr] 
$$
\begin{equation}\label{eq24}
\simeq 
T_0 \frac{5}{2\eta_0^{decay}}
\biggl(\frac{6}{5} - \frac{\Omega_0}{\omega_0}\biggr).
\end{equation}
As follows from Eqs.~(20), (23), and (24), $\varepsilon_F$ changes its sign
at $\Omega_0/\omega_0 =6/5$ both at weak and relatively strong pumping.
In the case under consideration, the population inversion is weak
($\varepsilon_F$ can be positive , but small) 
or is not realized ($\varepsilon_F < 0$).

Formally, the logarithmic factors in Eqs.~(22) and (23) diverge when 
$\Omega_0/\omega_0$ tends to $ 1 - \eta_0^{eq}/\eta_0^{decay}$. However,
these equations become invalid in the immediate vicinity
of the point $\Omega_0/\omega_0 =  1 - \eta_0^{eq}/\eta_0^{decay}$, 
because above we have assumed $\varepsilon_F/T < 1$ and put
$ \eta_0 =  \eta_0^{eq}$. The case of smaller  $\Omega_0/\omega_0$ is considered in the next subsection.

\subsection{Strong pumping, small $\Omega_0/\omega_0$ (cooling regime).}

The case  $\Omega_0/\omega_0$ is small, is much more 
interesting because rather large values of $\varepsilon_F$ and 
$\varepsilon_F/T$ can be achieved at sufficiently large $I/I_0$.
In such an instance, one can expect that
$\varepsilon_F \gg T$ and $\eta_0 \simeq (\hbar\omega_0/\varepsilon_F)/6$. 
Hence, Eq.~(14) can be presented as
\begin{widetext}
\begin{equation}\label{eq25}
T \simeq \frac{T_0}{1 - \displaystyle\frac{T_0}{\hbar\omega_0}\,
\ln\biggl\{1 +
\biggl[\eta_0^{decay} + \frac{1}{6}\biggl(\frac{\hbar\omega_0}{\varepsilon_F}\biggr)^2\biggl(\frac{\Omega_0}{\omega_0} - 1\biggr)
\biggr]\frac{I}{I_0}\biggr\}},
\end{equation}
\begin{equation}\label{eq26}
\varepsilon_F \simeq \frac{T_0}{2}\,
\frac{\ln \biggl\{\frac{\displaystyle 1 + 
\displaystyle (\eta_0^{decay} + 1)
\frac{I}{I_0}}
{\displaystyle 1 + 
\displaystyle\biggl[\eta_0^{decay} +  
\frac{1}{6}\biggl(\frac{\hbar\omega_0}{\varepsilon_F}\biggr)^2\biggl(\frac{\Omega_0}{\omega_0} - 1\biggr)\biggr]
\frac{I}{I_0}}\biggr\}}{1  - \displaystyle\frac{T_0}{\hbar\omega_0}\,
\ln\biggl\{1 +
\biggl[\eta_0^{decay} + \frac{1}{6}\biggl(\frac{\hbar\omega_0}{\varepsilon_F}\biggr)^2\biggl(\frac{\Omega_0}{\omega_0} - 1\biggr)
\biggr]\frac{I}{I_0}\biggr\}}.
\end{equation}
\end{widetext}

Assuming that $\varepsilon_F \gg T$, one can  approximately solve
 Eqs.~(25) and (26)
by iterations, considering the smallness of $\exp(-\varepsilon_F/T)$,
 and arrive at
\begin{equation}\label{eq27}
T \simeq T_0\biggl[1 - 2\sqrt{\frac{\biggl(1  - \displaystyle\frac{\Omega_0}{\omega_0}\biggr)\frac{I}{I_0}}
{6\biggl(\displaystyle 1 + \eta_0^{decay}\frac{I}{I_0}\biggr)}}\biggr],
\end{equation}
\begin{equation}\label{eq28}
\varepsilon_F \simeq \hbar\omega_0
\sqrt{\frac{\biggl(1  - \displaystyle\frac{\Omega_0}{\omega_0}\biggr)\frac{I}{I_0}}
{6\biggl(\displaystyle 1 + \eta_0^{decay}\frac{I}{I_0}\biggr)}}.
\end{equation}
If  $I/I_0  \gg 1/\eta_0^{decay}$, from Eqs.~(27) and (28)
for $T$ and $\varepsilon_F$   we obtain their saturation values:
$T^{sat} \simeq T_0[1 - 2\sqrt{(1  - \Omega_0/\omega_0)/6\eta_0^{decay}}]$
and $\varepsilon_F^{sat} \simeq \hbar\omega_0
\sqrt{(1  - \Omega_0/\omega_0)/6\eta_0^{decay}}$ .
The corrections to Eqs.~(27) and (28) are exponentially small when $\Omega_0/\omega_0$ is not too close to unity.
Due to the above assumptions,
Eqs.~(27) and (28)  are valid until $2\varepsilon_F < \hbar\omega_0$, that yields
$$
\biggl(\frac{2T_0}{3\hbar\omega_0}\biggr)^2 \ll 
\frac{2(1 - \Omega_0/\omega_0)I/I_0}{3(1 + \eta_0^{decay})I/I_0} < 1.
$$
Considering that $(2T_0/3\hbar\omega_0)^2 \simeq 4/576 \ll 1$,
these equations are valid in rather wide range of pumping intensities,
(not too small), except the case when  $(1 -\Omega_0/\omega_0)$ 
is very close to
zero (this is the special case considered above).
Some complications can arise in a non-realistic case of very small $\eta_0^{decay}$.
Using Eqs.~(27) and (28), one can find the condition when $\varepsilon_F \gg T$
that was assumed in deriving Eqs.~(25) and (26) and their consequences
(see also Eq.~(5)):
$$
\frac{\Omega_0}{\omega_0} < 1 - 6\eta_0^{decay}\biggl(\frac{T_0}{\hbar\omega_0}\biggr)^2.
$$
At room temperature, the latter inequality yields
 $\Omega_0/\omega_0 < 1 -  0.1\eta_0^{decay}$.
This implies that  large $\varepsilon_F$ and $T < T_0$ can be achieved if 
$\eta_0^{decay} < 10$ and the ratio  $\Omega_0/\omega_0$ is chosen properly.

As it was mentioned above,
the effect of cooling under consideration is attributed to the effective photogeneration of low-energy electrons and holes.

Setting $T_0 = 300$~K, 
$\hbar\Omega = 900$~meV, we obtain $\hbar\Omega_0 \simeq 0.1$~eV and
$\Omega_0/\omega_0 \simeq 0.5$. For this case, assuming $\eta_0^{decay} = 1 - 3$,
we obtain $T^{sat} \simeq 127 - 200$~K and
 $\varepsilon_F^{sat} \simeq (33 -  58)$~meV. Hence, $\exp(\varepsilon_F^{sat}/2T)
\simeq 2.7 - 15.5$. These latter two 
values are markedly larger than those obtained
in Subsection~A for $\Omega_0/\omega_0 = 1$.
If $\eta_0^{decay} < 1$, the effect of cooling is more pronounced
and the quantities  $\varepsilon_F^{sat}$ and  $\exp(\varepsilon_F^{sat}/2T)$
can be fairly large.
At the  parameters  chosen for the above estimates, conditions~(5) are satisfied.

It is instructive that in the cooling regime in question, the optical phonon
system is heated. The effective optical phonon temperature $\Theta$
in this regime
is given by the same equation as in other regimes under consideration, i.e. 
by Eq.~(18), which  yields $\Theta \geq T_0$.

\begin{figure}[t]
\begin{center}
\includegraphics[width=7.0cm]{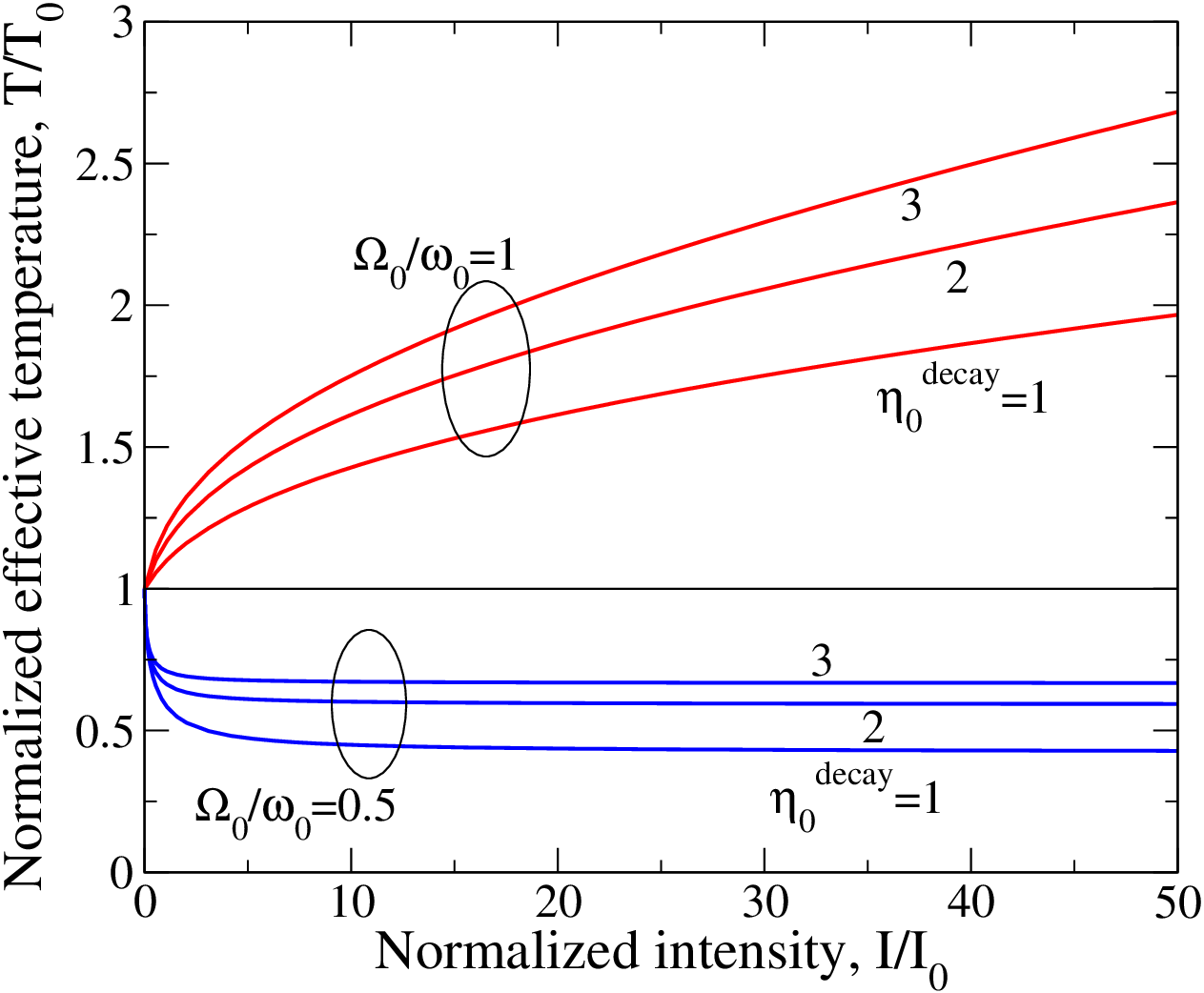}
\end{center}
\addvspace{-0 cm}\caption{ Normalized carrier effective temperature $T/T_0$ 
versus  
normalized pumping intensity $I/I_0$ for different
$\eta_0^{decay}$ and  $\Omega_0/\omega_0$.
}  
\end{figure}
\begin{figure}[t]
\begin{center}
\includegraphics[width=7.0cm]{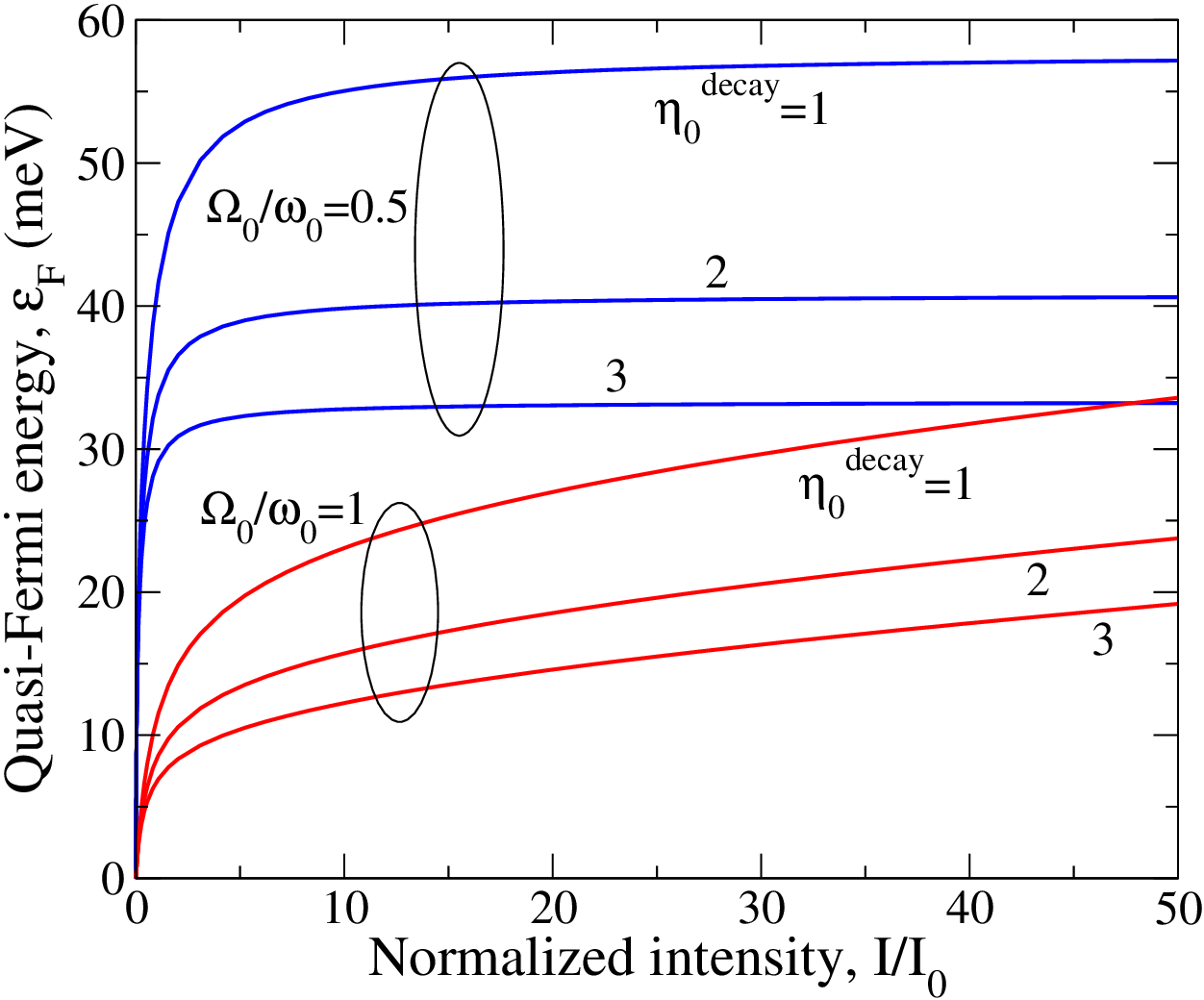}
\end{center}
\addvspace{-0 cm}\caption{Quasi-Fermi energy $\varepsilon_F$ 
versus  
normalized pumping intensity $I/I_0$ for different
$\eta_0^{decay}$ and  $\Omega_0/\omega_0$.
}  
\end{figure}

\section{Calculation of  the GL
dynamic conductivity}

The attenuation and amplification of electromagnetic waves
and surface plasmons 
propagating along the GL plane is determined by the
the real part of the GL dynamic ac conductivity at the signal frequency $\omega$ (photon or plasmon frequency)
Re~$\sigma_{\omega}$.
The latter includes
the contributions of the interband  and  intraband transitions:
\begin{equation}\label{eq29}
{\rm Re}~\sigma_{\omega} =
{\rm Re}\sigma_{\omega}^{inter} +  {\rm Re}~\sigma_{\omega}^{intra}.
\end{equation}
For the multiple-GL structures~(see, for instance, Refs.~\cite{9,34}), the contributions of all GL to
${\rm Re}~\sigma_{\omega}$ should be summarized~\cite{11,12}.

In the situation under consideration,
generalizing the pertinent formula from Ref.~\cite{6} 
(see, for instance, Refs.~\cite{2,11}) for 
the photogenerated nonequilibrium  electron hole plasma
with the equal non-zero quasi-Fermi energies of electrons and holes,
one can obtain:  
\begin{equation}\label{eq30}
{\rm Re }\sigma_{\omega}^{inter}  = \displaystyle\biggl(\frac{e^2}{4\hbar}\biggr)\tanh\biggl(\frac{\hbar\omega - 2\varepsilon_F}{4T}\biggr).
\end{equation}
Here, $e = |e|$ is the electron charge.
At $\omega \gg \tau^{-1}$, where $\tau$ is the momentum relaxation time,
The intraband contribution, which actually corresponds to the Drude
absorption by electrons and holes, can be obtained from
the Boltzmann equation~\cite{2,11,35}:
\begin{equation}\label{eq31}
{\rm Re }\sigma_{\omega}^{intra} = \frac{2e^2}{\pi\hbar^2\omega^2}
\int_0^{\infty}
 \displaystyle\frac{d\varepsilon\varepsilon}{\tau}
\frac{d}{d\varepsilon}\biggl\{-\frac{1}{\exp[(\varepsilon - \varepsilon_F)/T] + 1}\biggr\}.
\end{equation}
As $T_0$ and $T$ are small in comparison
with $\hbar\omega_0$, the  electron and hole momentum relaxation is
associated with different scattering mechanisms,
in particular, with the scattering on short-range and long-range disorder
and acoustic phonons. In this case, $\tau^{-1} = \nu_0 (\varepsilon/T_0)$,
($\nu_0$ is the collision frequency of electron and holes with
the thermal equilibrium energy),
so that interpolating the integral in Eq.~(31) [see Eq.~(A10)] as a function of $\varepsilon$
and $T$, from
Eq.~(31) we obtain
\begin{equation}\label{eq32}
{\rm Re }\sigma_{\omega}^{intra} \simeq 
\frac{2e^2\nu_0}{\pi\hbar^2\omega^2}\frac{(\varepsilon_F^2 + \pi^2T^2/6)}{T_0}.
\end{equation}
When $\varepsilon_F \gg T$ from
Eqs.~(31) and (32)    we obtain
\begin{equation}\label{eq33}
{\rm Re }\sigma_{\omega}^{intra} \simeq 
\frac{2e^2\nu_0}{\pi\hbar^2\omega^2}\frac{\varepsilon_F^2}{T_0}.
\end{equation}
Hence, using Eqs.~(30) - (33), we arrive at

\begin{equation}\label{eq34}
{\rm Re}~\sigma_{\omega}  =
\displaystyle\biggl(\frac{e^2}{4\hbar}\biggr)
\biggl[\tanh\biggl(\frac{\hbar\omega - 2\varepsilon_F}{4T}\biggr) +
\frac{8\nu_0}{\pi\hbar\omega^2}\frac{(\varepsilon_F^2 + \pi^2T^2/6)}{T_0}
\biggr].
\end{equation}
One can see that at $\tau^{-1} \propto \varepsilon,
$ the Drude term given by Eqs.~(34)  and (35) is proportional to the 
density of the electron-hole plasma.

In the special case considered above when $\Omega_0/\omega_0 = 1$,
using Eqs.~(16), (17), and (34), we obtain

\begin{widetext}
$$
\displaystyle\frac{{\rm Re}~\sigma_{\omega}}{\sigma_0} =
\frac{\displaystyle\frac{\exp(\displaystyle\hbar\omega/2T_0)}
{(1 + \eta_0^{decay}I/I_0)^{\omega/2\omega_0}} -
\biggl(1 + \frac{\displaystyle I/I_0}
{\displaystyle 1 + 
\displaystyle\eta_0^{decay}I/I_0}\biggr)^{1/2}}
{\displaystyle\frac{\exp(\displaystyle\hbar\omega/2T_0)}
{(1 + \eta_0^{decay}I/I_0)^{\omega/2\omega_0}}
 + \biggl(1 + \frac{\displaystyle I/I_0}
{\displaystyle 1 + 
\displaystyle\eta_0^{decay}I/I_0}\biggr)^{1/2}}
$$
\begin{equation}\label{eq35}
+
\biggl(\frac{\omega_D}{\omega}\biggr)^2\biggl\{\frac{3}{2\pi^2}
\ln^2\biggl(1 + \frac{I/I_0}
{\displaystyle 1 + 
\displaystyle\eta_0^{decay}I/I_0}\biggr)
 +  
\displaystyle\biggl[ 1 -\frac{T_0}{\hbar\omega_0}\,\ln(1 + \eta_0^{decay}I/I_0)\biggr]^{-2}\biggr\}.
\end{equation}
Here $\sigma_0 = e^2/4\hbar$ and $\omega_D = \sqrt{4\pi\nu_0T_0/3\hbar}$.
In the THz and middle IR ranges, $\omega \ll 2\omega_0$ ($\omega/2\pi \ll 10^3$~THz),
Eq.~(35) can be reduced to the following:
$$
\frac{{\rm Re}~\sigma_{\omega}}{\sigma_0} =
\frac{\displaystyle\exp\biggl(\displaystyle\frac{\hbar\omega}{2T_0}\biggr)  -
\biggl(1 + \frac{\displaystyle I/I_0}
{\displaystyle 1 + 
\displaystyle\eta_0^{decay}I/I_0}\biggr)^{1/2}}
{\displaystyle\exp\biggl(\displaystyle\frac{\hbar\omega}{2T_0}\biggr)
 + \biggl(1 + \frac{\displaystyle I/I_0}
{\displaystyle 1 + 
\displaystyle\eta_0^{decay}I/I_0}\biggr)^{1/2}}
$$
\begin{equation}\label{eq36}
+
\biggl(\frac{\omega_D}{\omega}\biggr)^2\biggl\{\frac{3}{2\pi^2}
\ln^2\biggl(1 + \frac{I/I_0}
{\displaystyle 1 + 
\displaystyle\eta_0^{decay}I/I_0}\biggr)
+  
\displaystyle\biggl[ 1 -\frac{T_0}{\hbar\omega_0}\,\ln(1 + \eta_0^{decay}I/I_0)\biggr]^{-2}\biggr\}.
\end{equation}
\end{widetext}

Assuming $\nu_0 = (0.25 - 1.0)\times10^{12}$~s$^{-1}$
at $T_0 = 300$~K, we obtain 
 $f_D = \omega_D/2\pi \simeq (1 - 2)$~THz.
Considering that the momentum relaxation time in multiple-GLs at the temperatures
$T_0 \leq 50$~K can reach values about 20~ps~\cite{34},
the above values of $\nu_0$ at $T_0 =300$~K appears to be reasonable
(assuming a linear increase of $\nu_0$ with the lattice temperature $T_0$~\cite{35}).

When $\Omega_0/\omega_0$ is small,
using Eqs.~(27), (28), and (34),  we arrive at 

$$
\displaystyle\frac{{\rm Re}~\sigma_{\omega}}{\sigma_0} =
\frac{\exp\biggl[\displaystyle\frac{\hbar\omega }{2T_0}\frac{1}{(1 - 2Z)}\biggr] -
\exp\biggl[\frac{\hbar\omega_0}{T_0}\frac{Z}{(1 - 2Z)}\biggr]
}
{\exp\biggl[\displaystyle\frac{\hbar\omega }{2T_0}\frac{1}{(1 - 2Z) }\biggr] + \exp\biggl[\frac{\hbar\omega_0}{T_0}\frac{Z}{(1 - 2Z)}\biggr] }
$$ 
\begin{equation}\label{eq37}
+
\biggl(\frac{\omega_D}{\omega}\biggr)^2\biggl[\frac{6}{\pi^2}\biggl(\frac{\hbar\omega_0}{T_0}\biggr)^2Z^2 + (1 - 2Z)^2\biggr].
\end{equation}
Here
$Z = \sqrt{[(1 - \Omega_0/\omega_0)I/I_0]/(1 + \eta_0^{decay}I/I_0)}$.

\section{Discussion of the results.}

As follows from Sec.~V, 
to achieve the population inversion sufficiently strong for lasing,
the parameter $\Omega_0/\omega_0$ should  be  either equal to unity
(the special case) or  
smaller than unity (cooling regime).
Below we shall focus on this situation.

Figure~1 shows the dependences  of   the normalized effective temperature
$T/T_0$ 
on  the normalized pumping intensity $I/I_0$ calculated 
for different $\eta_0^{decay}$ 
using eq.~(15) for  $\Omega_0/\omega_0 = 1$ and Eq.~(27) for  
$\Omega_0/\omega_0 = 0.5$.
Figure~2 shows  the dependences  of  the quasi-Fermi energy $\varepsilon_F$
on  the normalized pumping intensity $I/I_0$ calculated 
for different $\eta_0^{decay}$ and  $\Omega_0/\omega_0$
using Eqs.~(16) and (28).

One can see from Figs.~1 and 2 that at $\Omega_0/\omega_0 < 1$, a pronounced cooling
of the electron-hole plasma takes place, and this cooling 
regime provides larger $\varepsilon_F$ for a given pumping intensity.
As also follows from Figs~1 and 2, the finiteness of the optical decay time
markedly affects the $T - I$ and $\varepsilon_F - I$ dependences.
Indeed, at $\Omega_0/\omega_0 = 1 $, the $T - I$ dependence becomes steeper 
when the parameter $\eta_0^{decay}$ increases (reinforcement of heating effect), 
whereas at  $\Omega_0/\omega_0 < 1$,
this dependence becomes less pronounced (weakening of cooling).
This is attributed to the optical phonon heating, which promotes the heating
of the electron-hole plasma at  $\Omega_0/\omega_0 = 1 $ and counteract
its cooling at  $\Omega_0/\omega_0 < 1$.

Figure~3  demonstrates the dependences of the real part of the  
GL dynamic conductivity
Re~$\sigma_{\omega}$ on the signal frequency $\omega/2\pi$ at different
normalized pumping intensities $I/I_0$  calculated for  
different $\eta_0^{decay}$  and $\nu_0$ using Eq.~(35) (assuming that $\Omega_0/\omega_0 = 1$). As seen from Fig.~3, at $\Omega_0/\omega_0 = 1$,
the dynamic conductivity Re~$\sigma_{\omega}$ can be negative
only if parameter $\eta_0^{decay} \ll 1$ (the optical phonon decay time is small),
i.e., if the heating of electrons, holes, and optical phonons is very weak.
However,  at $\Omega_0/\omega_0 < 1$ when electron-hole plasma is cooled,
Re~$\sigma_{\omega} < 0$ in  range of signal  frequencies
from $\omega/2\pi \simeq 2$~THz to 17~THz and more [see Figs.~4 and 5, obtained using Eq.~(37)].
This occurs at   not too low
pumping intensities ($I/I_0 = 5$) but even 
at rather large values of $\nu_0$ and  $\eta_0^{decay}$ when the cooling of
the electron-hole plasma    is accompanied by the heating of the optical phonon system.
\begin{figure}[t]
\begin{center}
\includegraphics[width=7.0cm]{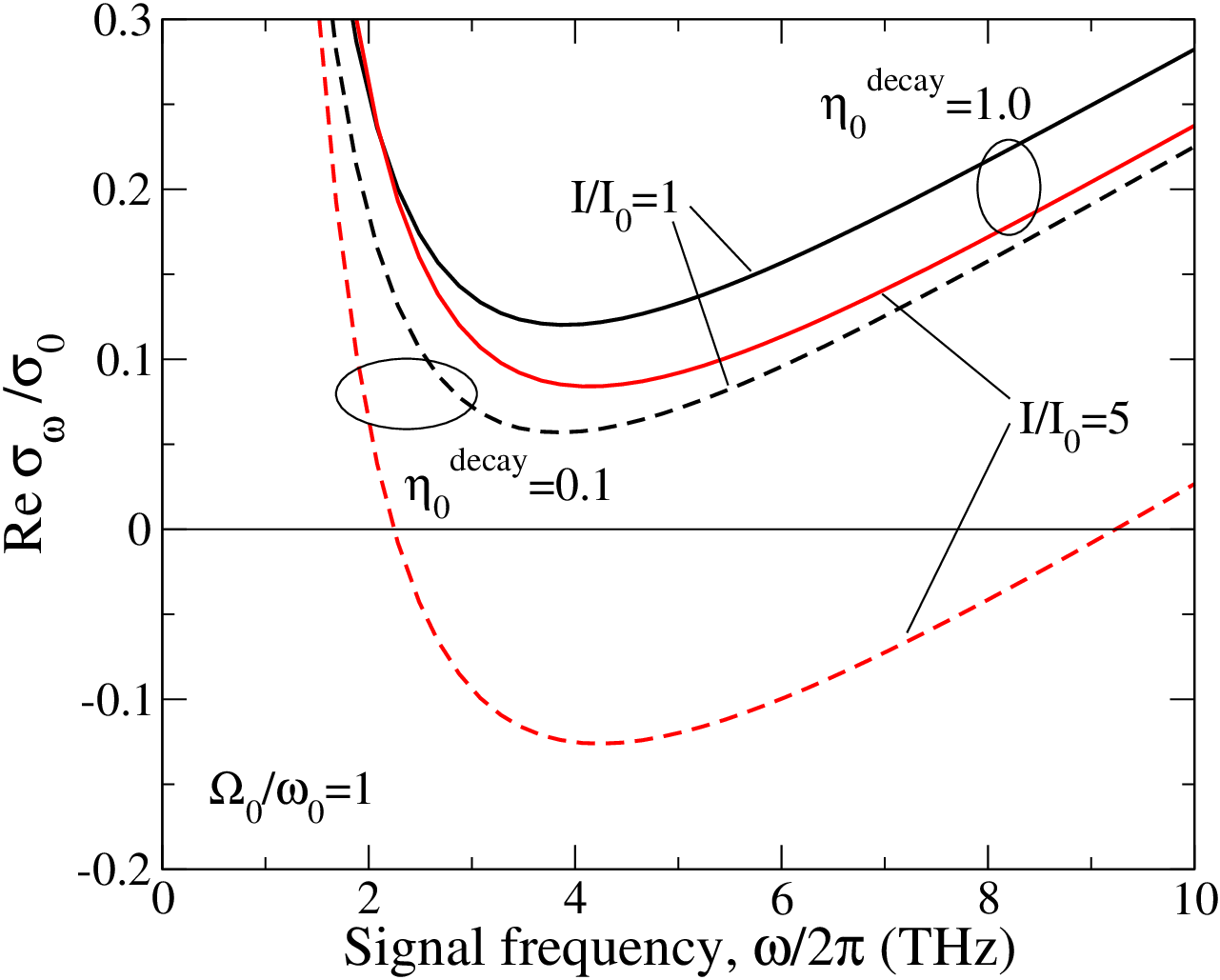}
\end{center}
\addvspace{-0 cm}\caption{Real part of the normalized
dynamic conductivity as a function of signal frequency at different  
normalized pumping intensities $I/I_0$ and different  $\eta_0^{decay}$
( $\Omega_0/ \omega_0 = 1$ and  $\nu_0 = 0.25\times 10^{12}$~s$^{-1}$).
}  
\end{figure}

Figures~6 and 7 calculated using Eq.~(37) show  the real part of the  
GL dynamic conductivity
Re~$\sigma_{\omega}$ versus  the normalized pumping intensity $I/I_0$
calculated for given values of the signal frequency $\omega/2\pi = 2$~THz
(at $\nu_0 = 2.5\times 10^{11}$~s$^{-1}$) and  $\omega/2\pi = 3.5$~THz.
(at $\nu_0 = 1\times 10^{12}$~s$^{-1}$). In both cases, Re~$\sigma_{\omega}$ 
is negative when $I/I_0$ exceeds certain threshold
value. The latter is somewhat smaller in GL with smaller $\nu_0$.
It is interesting that in the cases $\nu_0^{decay} = 1$, an increase in $I/I_0$
leads to smaller $|{\rm Re}~\sigma_{\omega}|$  in the range of  large $I/I_0$,
in contrast to the cases  $\nu_0^{decay} = 2$ and  $\nu_0^{decay} = 3$,
This can be explained by a faster increase in $\varepsilon_F$ with increasing
 $I/I_0$ at smaller $\nu_0^{decay}$ leading to a faster increase in the 
Drude absorption.

To estimate $I_0$, we use the data for 
 $\Sigma_0/\tau_0^{inter}$ at $T_0 = 300$~K 
from Ref.~\cite{15} (see Appendix) and set  $\hbar\Omega_0 = 100$~meV, so that 
($\Omega_0/\omega_0 \simeq 0.5$).
 and setting  $\hbar\Omega_0 = 100$~meV.
The chosen value of $\hbar\Omega_0$ can correspond to
the optical pumping photon energy $\hbar\Omega = 900$~meV
provided the cascade emission of $K = 2$ optical phonons or  
to $\hbar\Omega =  \hbar\Omega_0 = 100$~meV (pumping by CO$_2$ laser).
In such cases, one obtains 
$I_0 \simeq 9,66\times10^{21}$~1/cm$^{2}$s
and $I_0 \simeq 8.69\times10^{22}$~1/cm$^{2}$s, respectively.
In both cases, $S_0 = \hbar\Omega\,I_0 \simeq 1.39\times 10^4$ W/cm$^{2}$.

As demonstrated, lowering of the ration $\Omega_0/\omega_0$  
is beneficial for achieving the population inversion and 
negative dynamic conductivity,
because such lowering results in weakening of the electron-hole plasma heating and
even in the transition to its cooling.
Moreover,  as follows from Eq.~(23) and (24), 
at elevated $\Omega_0/\omega_0$, the value of the
quasi-Fermi
energy $\varepsilon_F$   can become negative that corresponds to nondegenerate
electron-hole plasma ($f < 0.5$). 
In this regard, in the case of optical pumping with relatively
high photon energy,
the suppression of the pair carrier interactions might be indispensable
for the realization of strong population inversion and lasing.
This, possibly, can be done using the GL structures with high-k  substrates. 

\begin{figure}[t]
\begin{center}
\includegraphics[width=7.0cm]{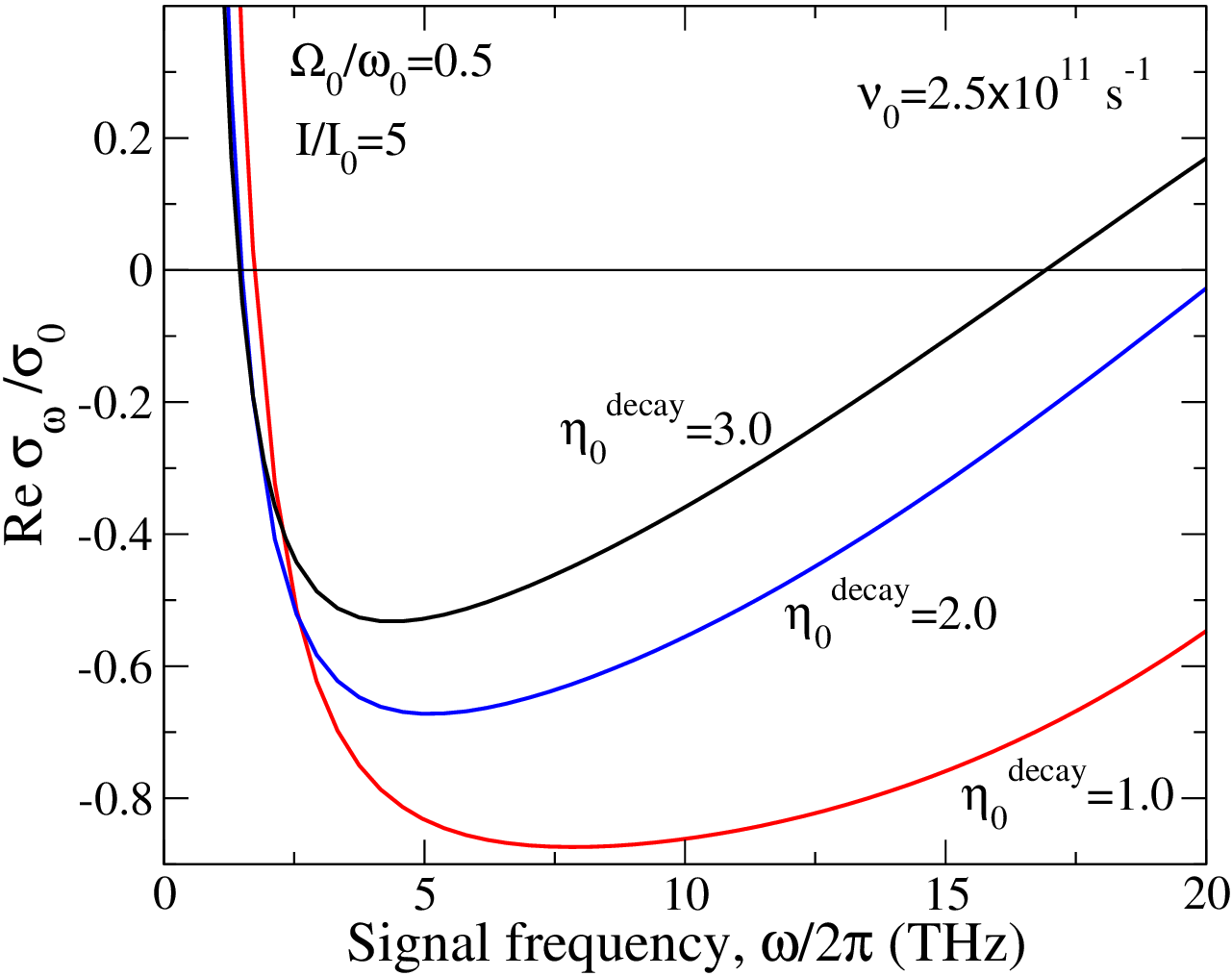}
\end{center}
\addvspace{-0 cm}\caption{Real part of the normalized
dynamic conductivity as a function of signal frequency at different  
normalized pumping intensities $I/I_0$ and different  $\eta_0^{decay}$
( $\Omega_0/ \omega_0 = 0.5$ and  $\nu_0 = 2.5\times 10^{11}$~s$^{-1}$).
}  
\end{figure}

\begin{figure}[t]
\begin{center}
\includegraphics[width=7.0cm]{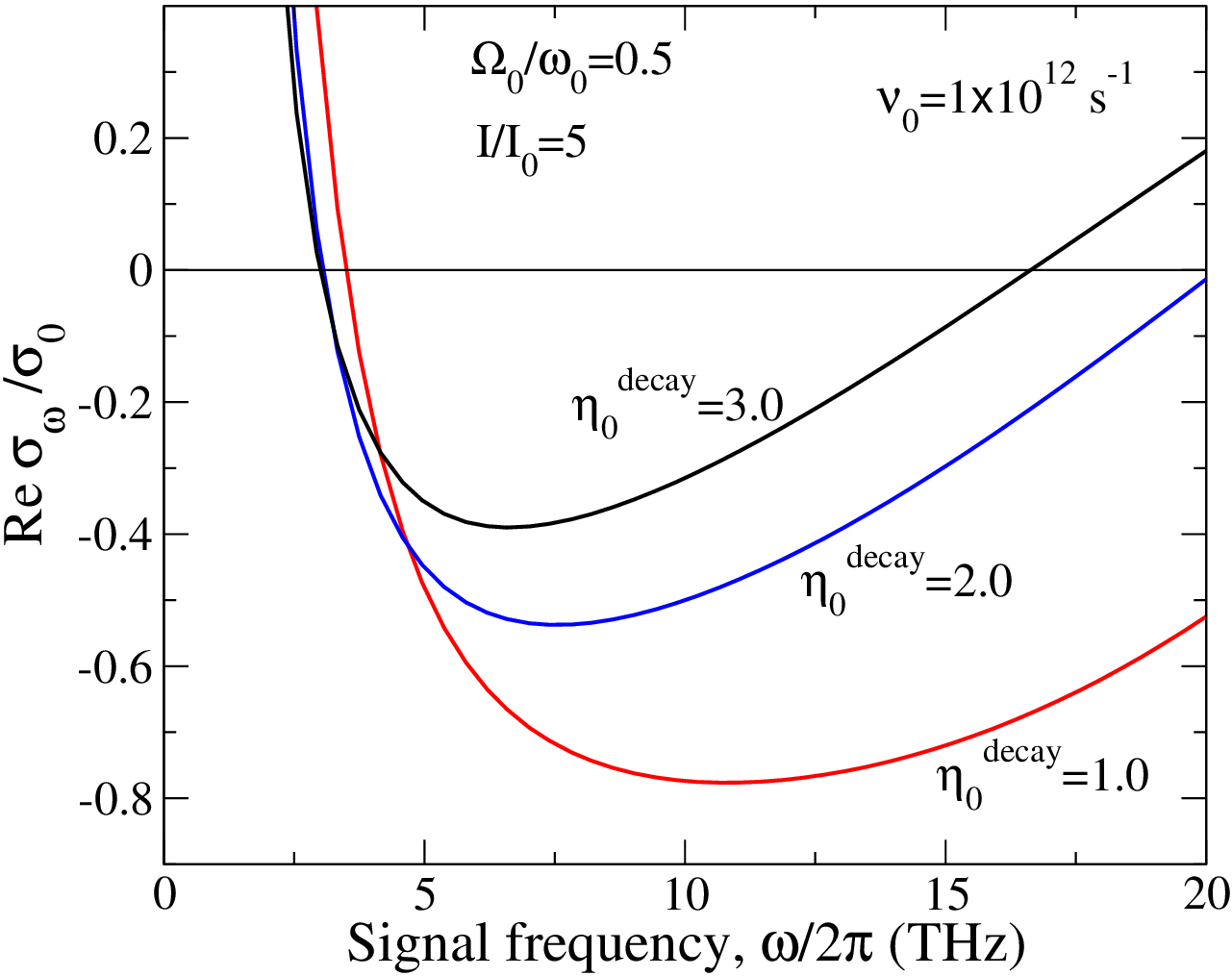}
\end{center}
\addvspace{-0 cm}\caption{The same as in Fig.~4 but for  
$\nu_0 = 1.0\times 10^{12}$~s$^{-1}$.
}  
\end{figure}
\begin{figure}[t]
\begin{center}
\includegraphics[width=7.0cm]{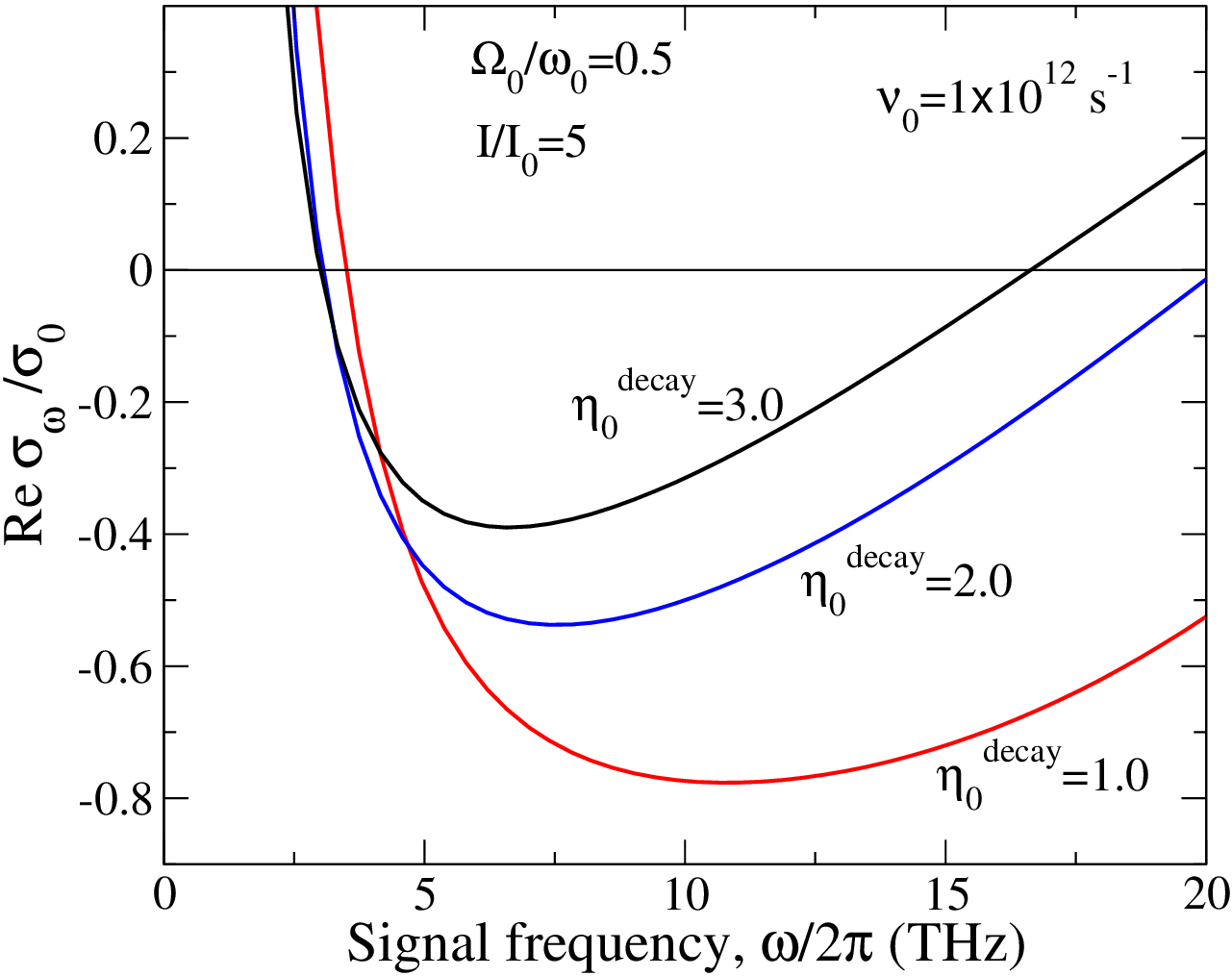}
\end{center}
\addvspace{-0 cm}\caption{Real part of the normalized
dynamic conductivity as a function of   
normalized pumping intensities $I/I_0$ for $\omega/2\pi = 2$~THz
and  different  $\eta_0^{decay}$
( $\Omega_0/ \omega_0 = 0.5$ and  $\nu_0 = 2.5\times 10^{11}$~s$^{-1}$).
}  
\end{figure}

\begin{figure}[t]
\begin{center}
\includegraphics[width=7.0cm]{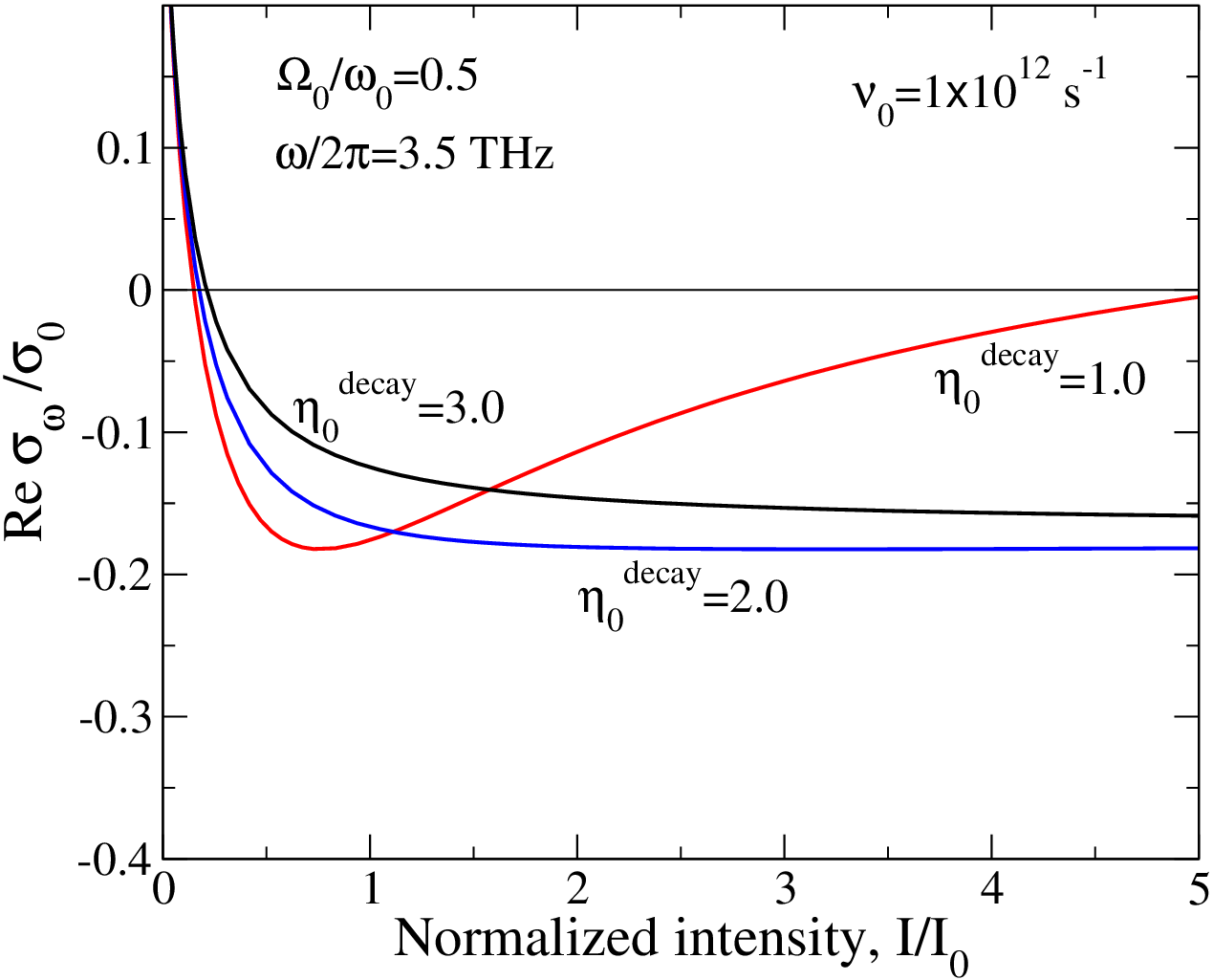}
\end{center}
\addvspace{-0 cm}\caption{The same as in Fig.~6 but for  for $\omega/2\pi = 3.5$~THz
and $\nu_0 = 1.0\times 10^{12}$~s$^{-1}$.
}  
\end{figure}
Instead of the  optical pumping under consideration,
the electron-hole cooling and the realization of strong population inversion
can be achieved by the injection pumping (in GL structures with
p-n or p-i-n junctions) considered previously~\cite{3}.
In such a case, one can put $\Omega_0 = eV/\hbar \ll \omega_0$, if 
 the applied bias voltage
 $V$ is not large ($V \ll 200$mV).

Above we assumed for simplicity  that there is only one type of optical phonons
with $\hbar\omega_0 \simeq 200$~meV. Actually, the processes involving
the 
 optical phonons with $\omega_1 = \omega_0)$ and $\omega_2 = \omega_0 - \Delta\omega_0$
are important~\cite{16} ($\Delta\omega_0/\omega_0 \simeq 0.18$). This adds complexity 
to the pattern of the electron and hole relaxation affecting the value of parameter
$\Omega_0$ and

 Although one can expect that
due to a smallness of 
$\Delta\omega_0/\omega_0$, this effect is not essential
and can be accounted for by a proper correction of $\Omega_0$.  .  
More complicating  factor can be the interaction of electrons and holes with
optical phonons in the substrate. These phonons, having relatively low energy, can,
in principle, markedly contribute to the recombination and energy relaxation.
Perhaps in multiple GL-structures the effect substrate phonons is not pronounced.
The pertinent generalization of our treatment is going to be considered elsewhere.

Lowering of the lattice temperature $T_0$ and, consequently,
the effective temperature $T$  results a weakening of the efficiency
of optical phonon assisted processes, particularly, in a decrease in  the recombination rate [see Eqs.~(6) and (7)]. Apart from this, the Drude absorption also 
becomes smaller due to a decrease
in $\nu_0$ [see Eq.~(34)]. Therefore, although, as demonstrated above,
the THz and IR lasing using the population   
inversion in graphene can be  realized even at room temperatures,
at lower temperatures, it can be achieved easier.
However, at lower temperatures, the optical phonon assisted processes 
in GLs can give
way to 
other recombination, generation,  and energy relaxation mechanisms, for instance,
the radiative recombination and generation and the energy relaxation 
on acoustic phonons~\cite{13,14,36}. 
To  estimate the temperature range, in which the above simplified
"optical phonon" model
is sufficient, we assume that the characteristic time of radiative recombination
and that associated with the optical phonon emission are about
(10 - 20)~ns and 0.5~ps, respectively. In this case, one can obtain
$T_0 > 220 - 230$~K, so that our model is valid at room temperatures and slightly
lower.



\section{Conclusions}

We have considered the characteristics of electron-hole plasma in optically
pumped GL at elevated (room)  temperatures (its quasi-Fermi energy,
effective temperature, and dynamic conductivity) and shown that:\\
(1) The interband and intraband processes of emission and absorption of optical phonons play a crucial role in the characteristics of optically pumped electron-hole plasma in graphene at elevated temperatures;\\
(2) The electron-hole plasma in optically pumped graphene can be  both
heated and cooled depending on  parameter
 $\Omega_0/\omega_0$, while the optical phonon system
is always heated;\\
(3) The accumulation of nonequilibrium optical phonons (their heating) due to
finiteness of their decay time leads to a slower dependence
 of the electron and hole quasi-Fermi energy on the  optical pumping 
intensity;\\
(4) The effects in question can markedly influence the achievement of
the negative dynamic conductivity in optically pumped
GLs associated with the population inversion and, hence, the realization
of THz or IR lasing in the GL structures at room temperatures.
The latter requires a careful choice of parameter  $\Omega_0/\omega_0$
and minimization of parameter $\eta_0^{decay}$.

\section{Acknowledgment}
The work was supported by  the Japan Science and 
Technology Agency, CREST,  Japan.

\section*{Appendix}
\setcounter{equation}{0}
\renewcommand{\theequation} {A\arabic{equation}}


 For the term $R_0^{inter}$ in Eq.~(6), which describes the
electron-hole recombination and generation processes assisted by optical phonons,
 one can use the following simplified
formula:

\begin{widetext}
$$
R^{inter} = \frac{G_0}{T_0^3}
\int_{0}^{\hbar\omega_0}
\frac{\displaystyle\,d\varepsilon\varepsilon(\hbar\omega_0 - \varepsilon)}
{\displaystyle\biggl[ 1 + \exp\biggl(\frac{\varepsilon - \varepsilon_F}{T}
\biggr)\biggr] \biggl[ 1 + \exp\biggl(\frac{\hbar\omega_0 - \varepsilon - 
\varepsilon_F}{T}\biggr)\biggr]}\biggl[{\cal N}_0 + 1 - 
\exp\biggl(\frac{\hbar\omega_0 - 2\varepsilon_F}{T}\biggr){\cal N}_0\biggr]
$$
$$
= \frac{G_0(\hbar\omega_0)^2T}{6T_0^3}\biggl[({\cal N}_0 + 1)
\exp\biggl(\frac{2\varepsilon_F -\hbar\omega_0}{T}\biggr) - 
{\cal N}_0\biggr]
\int_{\exp(- \varepsilon_F/T)}^{\exp[(\hbar\omega_0 - \varepsilon_F)/T]}
\frac{\displaystyle\,du}
{\displaystyle(1 + u)
[ 1 + ue^{(\varepsilon_F - \hbar\omega_0)/T}]}
$$
\begin{equation}\label{eqA1}
= \frac{G_0(\hbar\omega_0)^2T}{6T_0^3}\biggl[({\cal N}_0 + 1)
\exp\biggl(\frac{2\varepsilon_F -\hbar\omega_0}{T}\biggr) - 
{\cal N}_0\biggr]
\frac{\displaystyle\exp\biggl(\frac{\hbar\omega_0 - 2\varepsilon_F}{T}\biggr)}
{\displaystyle\biggl[\exp\biggl(\frac{\hbar\omega_0 - 2\varepsilon_F}{T}\biggr) - 1\biggr]}
\ln\biggl\{\exp\biggl(-\frac{\hbar\omega_0}{T}\biggr)
\biggl[\frac{\displaystyle\\exp\biggl(\frac{\hbar\omega_0 - \varepsilon_F}{T}\biggr) + 1}
{\displaystyle\\exp\biggl( - \frac{\varepsilon_F}{T}\biggr) + 1}\biggr]^2
\biggr\}.
\end{equation}
\end{widetext}
Equation~(A1) can be markedly simplified
in the most interesting situations 
when   $\hbar\omega_0> 2\varepsilon_F,  T$:

\begin{equation}\label{eqA2}
R^{inter} \simeq  \frac{\Sigma_0}{\tau_0^{inter}}\biggl[({\cal N}_0 + 1)
\exp\biggl(\frac{2\varepsilon_F -\hbar\omega_0}{T}\biggr) - 
{\cal N}_0\biggr] 
\end{equation}
with 
 \begin{equation}\label{eqA3}
\frac{\Sigma_0}{\tau_0^{inter}} = \frac{G_0(\hbar\omega_0)^3}{6T_0^3}.
\end{equation}
Here we have neglected the terms in the pre-exponential factor with the ration 
$\varepsilon_F/\hbar\omega_0$ and introduced the rate of the thermal generation
of the electron-hole pairs  due to absorption of optical phonons in equilibrium $G_{0}^{eq} = G_0(\hbar\omega_0)^3/6T_0^3
\exp(- \hbar\omega_0)/T) =
(\Sigma_0/\tau_0^{inter})
\exp(- \hbar\omega_0/T)$. The latter 
was  estimated in Ref.~\cite{16}: $G_0^{eq}\simeq 10^{21}$~cm$^{-2}$s$^{-1}$
at $T_0 = $ 300K.

Equation~(A2) differs from those in Refs.~\cite{11,16} by 
the inclusion of the terms
with ${\cal N}_0$, which  correspond to the processes of stimulated 
emission
and absorption of optical phonons 
At  temperatures, 
$T \ll \hbar\omega_0 \simeq$~2300~K, when 
the number of equilibrium optical
phonons  ${\cal N}_0 =
{\cal N}_0^{eq} \ll 1$ (stimulated  emission
and absorption of optical phonons is negligible),
 eq.~(A2) reduces to that obtained 
in ref.~\cite{11}.


The term $R_0^{intra}$ in Eq.~(7) can be calculated as
\begin{widetext}
\begin{equation}\label{eqA4}
R_0^{intra} = \frac{2G_0}{T_0^3}
\int_0^{\infty}\frac{d\varepsilon\varepsilon(\varepsilon + \hbar\omega_0)
\displaystyle\biggl[{\cal N}_0 + 1\exp\biggl(-\frac{\hbar\omega_0}{T}\biggr) 
-  {\cal N}_0\biggr] }
{\biggl[1 + \displaystyle\exp\biggl(\frac{\varepsilon - \varepsilon_F}{T}\biggr)\biggr]
\displaystyle\biggl[
1 + \displaystyle\exp\biggl(- \frac{\varepsilon - \varepsilon_F + \hbar\omega_0}{T}\biggr)\biggr]} 
\simeq 
\frac{\Sigma_0}{\tau_0^{intra}}\biggl[({\cal N}_0 + 1)\exp\biggl(-\frac{\hbar\omega_0}{T}\biggr) 
- {\cal N}_0\biggr],
\end{equation}
\end{widetext}
where 
\begin{equation}\label{eqA5}
\frac{\Sigma_0}{\tau_0^{intra}} = 2G_0\biggl(\frac{T}{T_0}\biggr)^3\int_0^{\infty}\frac{duu(u + \hbar\omega_0/T)
\displaystyle }
{\biggl[1 + \displaystyle\exp(u - \varepsilon_F/T)
\biggr]}.
\end{equation}

If  $ |\varepsilon_F| <  T$, Eq.~(A4) yields
\begin{equation}\label{eqA6}
\frac{\Sigma_0}{\tau_0^{intra}} \simeq \frac{\pi^2G_0}{6}\biggl(\frac{T}{T_0}\biggr)^2
\biggl(\frac{\hbar\omega_0 + 2.19T}{T_0}\biggr).
\end{equation}
Comparing Eqs.~(A3) and (A6), for this case we obtain

\begin{equation}\label{eqA7}
\eta_0 =  \frac{\tau_0^{intra}}{\tau_0^{inter}}
= \frac{1}{\pi^2}\biggl(\frac{\hbar\omega_0}{T}\biggr)^2
\biggl(1 + 2.19\frac{T}{\hbar\omega_0}\biggr)^{-1}.
\end{equation}
If $\hbar\omega_0 >  \varepsilon_F \gg T$,
\begin{equation}\label{eqA8}
\frac{\Sigma_0}{\tau_0^{intra}} \simeq  G_0\frac{\hbar\omega_0\varepsilon_F^2(1 + 2\varepsilon_F/3\hbar\omega_0)}{T_0^3}\simeq  G_0\frac{\hbar\omega_0\varepsilon_F^2}
{T_0^3},
\end{equation}
 so that invoking Eq.~(A3), we obtain
\begin{equation}\label{eqA9}
\eta_0
 \simeq \frac{1}{6}\biggl(\frac{\hbar\omega_0}{\varepsilon_F}\biggr)^2.
\end{equation}

In a wide range of variations of $\varepsilon_F$ and $T$,
one can use the following interpolation:
\begin{equation}\label{eqA10}
\eta_0
=  \frac{\hbar^2\omega_0^2}{(6\varepsilon_F^2 + \pi^2T^2)}.
\end{equation}

\end{document}